\documentclass[10pt,twocolumn, journal]{IEEEtran}
\usepackage[T1]{fontenc}% optional T1 font encoding
\ifCLASSINFOpdf
\else
\fi
\usepackage{amsmath}
\usepackage{amsbsy}
\usepackage{amssymb}
\usepackage{amsthm}
\usepackage{graphics}
\usepackage{epsfig}
\usepackage{graphicx}
\usepackage{psfrag}
\usepackage{cite}
\usepackage{color}
\usepackage{cases}
\usepackage{bm}
\newtheorem{theorem}{Theorem}
\newtheorem{lemma}{Lemma}

\newtheorem{remark}{Remark}

\newtheorem{corollary}{Corollary}
\ifCLASSOPTIONcompsoc
\usepackage[caption=false,font=normalsize,labelfont=sf,textfont=sf]{subfig}
\else
\usepackage[subrefformat=parens,labelformat=parens,caption=false,font=footnotesize]{subfig}
\fi
\date{August 25, 2018}
\usepackage{url}
\IEEEoverridecommandlockouts
\begin{document}

\title{\LARGE Per-link Reliability and Rate Control: \\Two Facets of the SIR Meta Distribution}
\author{Sanket~S.~Kalamkar,~\IEEEmembership{Member,~IEEE,}
       and~Martin~Haenggi,~\IEEEmembership{Fellow,~IEEE}
\thanks{S. S. Kalamkar is with INRIA, Paris, 75012, France. M. Haenggi is with the Department of Electrical Engineering, University of Notre Dame, Notre Dame, IN, 46556 USA. (e-mail: $\lbrace$skalamka, mhaenggi$\rbrace$@nd.edu).}
\thanks{This work is supported by the US National Science Foundation (grant CCF 1525904).}
}
\maketitle

\begin{abstract}
The meta distribution (MD) of the signal-to-interference ratio (SIR) provides fine-grained reliability performance in wireless networks modeled by point processes. In particular, for an ergodic point process, the SIR MD yields the distribution of the per-link reliability for a target SIR. Here we reveal that the SIR MD has a second important application, which is rate control. Specifically, we calculate the distribution of the SIR threshold (equivalently, the distribution of the transmission rate) that guarantees {\em each} link a target reliability and show its connection to the distribution of the per-link reliability. This connection also permits an approximate calculation of the SIR MD when only partial (local) information about the underlying point process is available.
\end{abstract}

\section{Introduction}

When the wireless node locations are modeled by a point process, the meta distribution (MD) of the signal-to-interference ratio (SIR) provides refined network performance by calculating the per-link reliability~\cite{martin_meta_2016}. Specifically, for a point process $\Phi$ of transmitters, the conditional success probability (equivalently, the reliability) of a link is given by 
\begin{align}
P_{\rm{s}}(t) \triangleq \mathbb{P}(\mathsf{SIR} > t \mid \Phi), \quad t \in \mathbb{R}^{+},
\label{eq:cond_suc}
\end{align}
where $t$ is the SIR threshold. In \eqref{eq:cond_suc}, averaging is done over the fading and the channel access scheme. The SIR MD at the typical link is obtained by averaging over the point process as 
\begin{align}
\bar{F}_{P_{\rm{s}}}(t, x) = \bar{F}_{P_{\rm{s}}(t)}(x)\triangleq \mathbb{P}(P_{\rm{s}}(t) > x), \quad x \in [0, 1],
\label{eq:basic_MD}
\end{align}
where $x$ is the reliability threshold. In~\eqref{eq:basic_MD}, depending on the network model, we may need to use the reduced Palm measure instead of the probability measure given that an active transmitter is present at a prescribed location and the SIR is calculated at its associated receiver. For an ergodic point process model, the SIR MD can be interpreted as the fraction of links or users that achieve a reliability at least $x$ for a target SIR $t$ in each realization of $\Phi$.

A simple way to guarantee {\em each} link a target reliability in each realization of $\Phi$ is to {\em control} the (bandwidth-normalized) transmission rate $\mathcal{R}$, which depends on the SIR threshold through the spectral efficiency as $\mathcal{R} = \log(1+t)$.

In rate control, for a random $\Phi$, the SIR threshold for which a link achieves the target reliability is random. Let $T$ denote this random SIR threshold at the typical user. The complementary cumulative distribution function (ccdf) of $T$ for a target reliability of $\nu$ is denoted by
\begin{equation}
\bar{F}_{T(\nu)}(t) \triangleq \mathbb{P}(T(\nu) > t).
\end{equation} 
It determines the rate distribution via $\mathcal{R} = \log(1+T)$.

\subsubsection*{Related work}

The SIR MD is calculated for the spectrum sharing between the device-to-device (D2D) and cellular users in \cite{martin_d2d} and for millimeter-wave D2D networks in \cite{deng_mm}. The works in \cite{martin_meta_2016, martin_d2d, deng_mm} take an approach where each link is subjected to the same SIR threshold and calculate the fraction of links that satisfy a target reliability in each realization of $\Phi$. Using the tool of the SIR MD, \cite{soc_2017} calculates the \textit{spatial outage capacity}, which is the maximum density of concurrently active links that meet a target success probability. In~\cite{sanket_globecom17}, the approach of rate control is used to achieve a high reliability given only the knowledge of the nearest interferer's location. In~\cite{weber_wiopt_17}, given the different levels of knowledge about the point process, the success probability of a transmission is predicted. This paper takes a dual approach to~\cite{martin_meta_2016} (and similar to \cite{sanket_globecom17}) in that the SIR threshold is adjusted to meet the target reliability.

\subsubsection*{Contributions} 
This paper reveals the two interpretations of the SIR MD---the per-link reliability distribution and the rate distribution. Specifically, it makes the following contributions:
\begin{itemize}
\item For a wireless network modeled by a stationary and ergodic point process, we prove that calculating the SIR MD as a function of the SIR threshold is equivalent to calculating the SIR threshold distribution such that each link is guaranteed a target reliability of $\nu$ (Thm.~\ref{thm:dual}), {\em i.e.},
\begin{align*}
\bar{F}_{P_{\rm{s}}(t)}(\nu) \equiv \bar{F}_{T(\nu)}(t).
\end{align*}
%This identity reveals the rate control as the second facet of the SIR MD---the performance of user percentiles being the first facet.
%The distribution of the SIR threshold ultimately results in the rate distribution, which gives insights into the rate-reliability trade-off.
\item For Poisson bipolar networks~\cite{baccelli_2006} with Rayleigh fading and the standard path loss model, we show that the rate control approach permits an approximation of the SIR MD given the knowledge of the locations of only a few nearest interferers (Thms.~\ref{thm:distr_T}, \ref{thm:ultrarel}, and \ref{thm:approx_T}).
\item We also show that the rate distribution facilitates the calculation of the throughput density.
\end{itemize}  \vspace*{-1mm}

\section{Duality in the Interpretation of the SIR MD}
\begin{theorem}
\label{thm:dual}
For any stationary and ergodic point process $\Phi$, given a target reliability $\nu$, 
\begin{align}
\bar{F}_{P_{\rm s}(t)}(\nu) \equiv \bar{F}_{T(\nu)}(t).
\end{align}
%where $\bar{F}_{T(\nu)}(t)$ is the ccdf of the SIR threshold $T$ that achieves a reliability equal to $\nu$.
\end{theorem}
\begin{IEEEproof}
For a target reliability $\nu$, the SIR MD as a function of the SIR threshold $t$ is\vspace*{-1mm}
\begin{align*}
\bar{F}_{P_{\rm s}(t)}(\nu) &=  \mathbb{P}(\mathbb{P}({\sf SIR} > t \mid \Phi)> \nu) \\
&\overset{(\mathrm a)}{=} \mathbb{P}((1- F_{{\sf SIR}\mid \Phi}(t)) > \nu) \\
&\overset{(\mathrm b)}{=} \mathbb{P}(t < F^{-1}_{{\sf SIR}\mid \Phi}(1-\nu)) \\
&\overset{(\mathrm c)}{=} \mathbb{P}(T(\nu) > t), \\
&= \bar{F}_{T(\nu)}(t),
\end{align*}\vspace*{-5mm}

\noindent where in step $\mathrm{(a)}$, $F_{{\sf SIR} \mid \Phi}(t)$ denotes the SIR cdf conditioned on $\Phi$. In step $\mathrm{(b)}$, $F^{-1}_{{\sf SIR}\mid \Phi}(1-\nu)$ denotes the inverse conditional cdf of the SIR. Since $F_{{\sf SIR}\mid \Phi}(t)$ is continuous and strictly increasing, $F^{-1}_{{\sf SIR}\mid \Phi}(1-\nu) = u$ is the unique number such that $F_{{\sf SIR}\mid \Phi}(u) = 1-\nu$. In step $\mathrm{(c)}$,\vspace*{-1mm} 
\begin{align}
T(\nu) \triangleq F^{-1}_{{\sf SIR}\mid \Phi}(1-\nu).
\label{eq:def_T}
\end{align}
Here $T(\nu)$ is random since $\Phi$ is random. Given $\Phi$, we have $T(\nu) = \theta$ such that 
\begin{align}
\theta &= F^{-1}_{{\sf SIR}\mid \Phi}(1-\nu),
\label{eq:def_inv_T}
\end{align}\vspace*{-5mm}

\noindent which results in\vspace*{-1mm} 
\begin{equation}
\nu = 1 - F_{{\sf SIR}\mid \Phi}(\theta) = \mathbb{P}({\sf SIR} > \theta \mid \Phi) \label{eq:def_inv_T2}.\vspace*{-1mm}
\end{equation}
From \eqref{eq:def_T}, \eqref{eq:def_inv_T}, and \eqref{eq:def_inv_T2}, it follows that $\bar{F}_{T(\nu)}(t)$ is the ccdf of the SIR threshold that achieves a reliability equal to $\nu$.
\end{IEEEproof}
\begin{corollary}
\label{cor:dual}
The distribution of the transmission rate $\mathcal{R}$ is\vspace*{-1mm}
\begin{align}
\bar{F}_{\mathcal{R}}(r) \equiv \bar{F}_{P_{\rm s}(e^r -1)}(\nu).
\end{align}
\end{corollary}\vspace*{-1mm}
\begin{IEEEproof}
The rate distribution follows from $\mathcal{R} = \log(1 + T)$ and Thm.~\ref{thm:dual}.
\end{IEEEproof}

Thm.~\ref{thm:dual} and Cor.~\ref{cor:dual} reveal two facets of the SIR MD:
\begin{itemize}
\item[1.] For fixed SIR threshold $t = \theta$, $\bar{F}_{P_{\rm s}}(\theta, x)$ is the per-link reliability distribution. 
\item[2.] For fixed reliability threshold $x = \nu$, $\bar{F}_{P_{\rm s}}(t, \nu)$ is the distribution of the conditional SIR threshold (equivalently, rate distribution).
\end{itemize}
Intuitively, Thm.~\ref{thm:dual} can be explained as follows. For a given SIR threshold $t$ and target reliability $\nu$, in a realization of the network, let us assume that the typical link{\footnote{We add a link whose receiver is at the origin. Under the expectation over $\Phi$, this link becomes the typical link. In a realization of $\Phi$, however, with a slight abuse of terminology, we term this link the typical link even before taking the expectation.}} achieves a reliability of $x$. If $x > \nu$, the SIR threshold can be increased until $x = \nu$. Contrary, if $x < \nu$, the SIR threshold needs to be decreased such that $x = \nu$. Hence the probability that the typical link achieves a reliability of $x > \nu$ is the same as the probability that the SIR threshold for which the typical link achieves a reliability of exactly $\nu$ is greater than $t$.
% For a stationary and ergodic point process, Thm.~\ref{thm:dual} further tells us that, in each realization of the network, the fraction of links that achieve a reliability at least $\nu$ for a given SIR threshold $t$ is equal to the fraction of links for which the SIR threshold is set to a value above $t$ while each link achieving the reliability of $\nu$.

% For a given SIR threshold, since the SIR MD is the probability that the typical link achieves a reliability of $x > \nu$, it is the same as the distribution of the SIR threshold for which the typical link achieves exactly the target reliability.
% For a stationary and ergodic point process, Thm.~\ref{thm:dual} further tells us that, in each realization of the network, the fraction of links that achieve a reliability at least $\nu$ for a given SIR threshold $t$ is equal to the fraction of links for which the SIR threshold is set to a value at least $t$ while each link achieving the reliability of $\nu$.

\vspace*{-1mm}

\section{The Case of Poisson Bipolar Networks}
In this section, we study the rate control performance in Poisson bipolar networks~\cite{baccelli_2006}, which can be used to model infrastructureless wireless networks such as ad hoc, D2D, M2M, and V2V networks.\vspace*{-2mm}

\subsection{Network Model}
\label{sec:net_mod}

The transmitter locations follow a homogeneous Poisson point process (PPP) $\Phi$ of density $\lambda$. Each transmitter has an associated receiver at a distance of $R$ in a uniformly random direction. Each transmitter transmits at unit power. We assume Rayleigh fading with the channel power gain distributed as the exponential random variable with mean $1$. A transmission is subject to the path loss as $r^{-\alpha}$, where $r$ is the distance and $\alpha > 2$ is the path loss exponent. We focus on the interference-limited case. We add a transmitter at the location $(R,0)$ and its receiver at the origin $o$. The link between this transmitter-receiver pair becomes the {\em typical link} under the expectation over $\Phi$.

\subsection{Per-link Reliability and Rate Control}

\begin{figure}\vspace*{-3mm}
  \centering
  \captionsetup{justification=centering}
  \subfloat[{Deterministic SIR threshold $\theta = 1$,\protect\\ $P_{\rm{s}}(\theta) \triangleq \mathbb{P}(\mathsf{SIR} > \theta | \Phi)$}\label{fig:ps_dist}]{\includegraphics[scale=0.15]{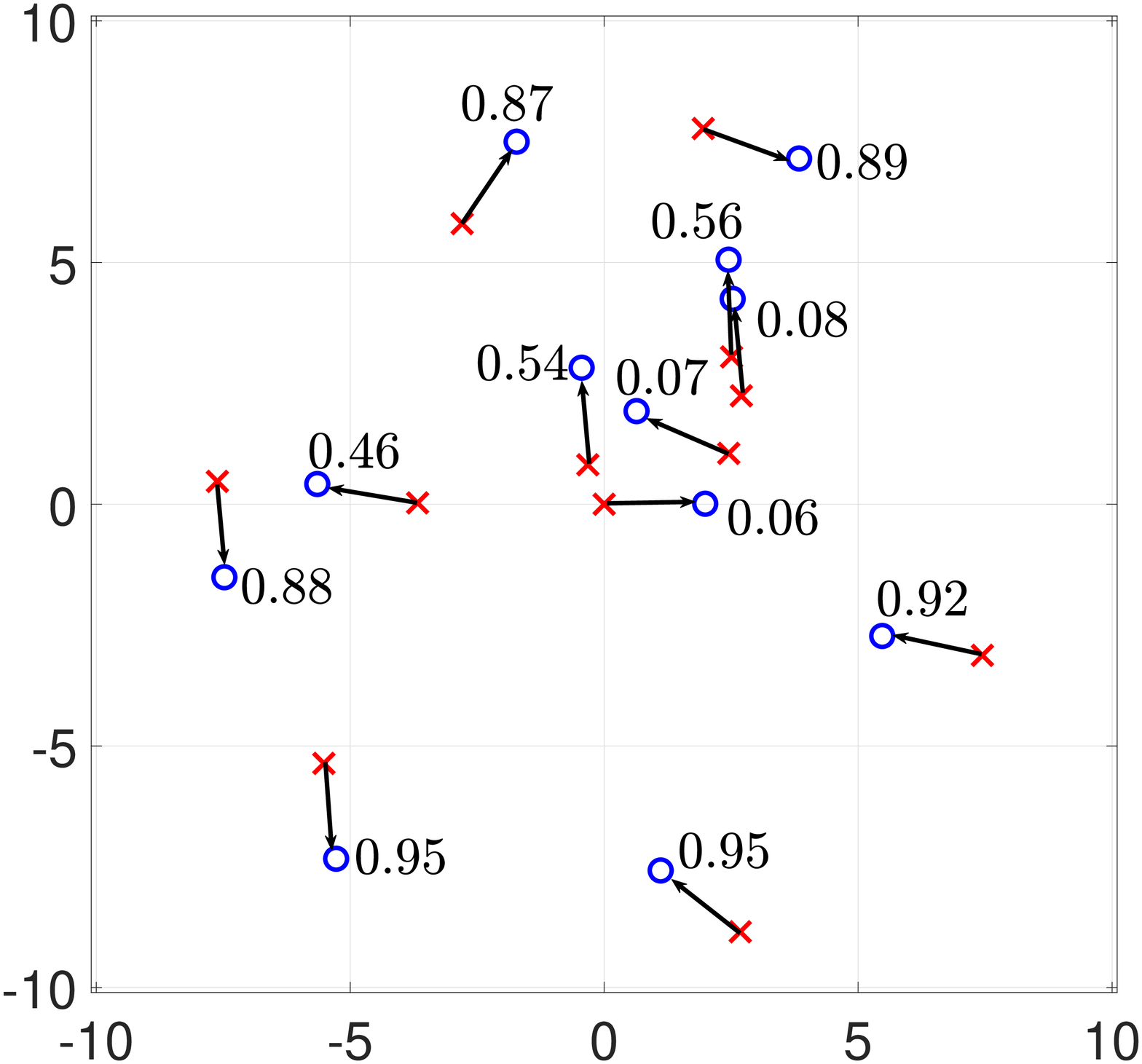}}
  \subfloat[Random SIR threshold $T$, $\nu= 0.9$,\protect\\  $T = F_{{\sf SIR}\mid \Phi}^{-1}(1-\nu) = P_{\rm s}^{-1}(\nu)$\label{fig:theta_dist}]{\includegraphics[scale=0.15]{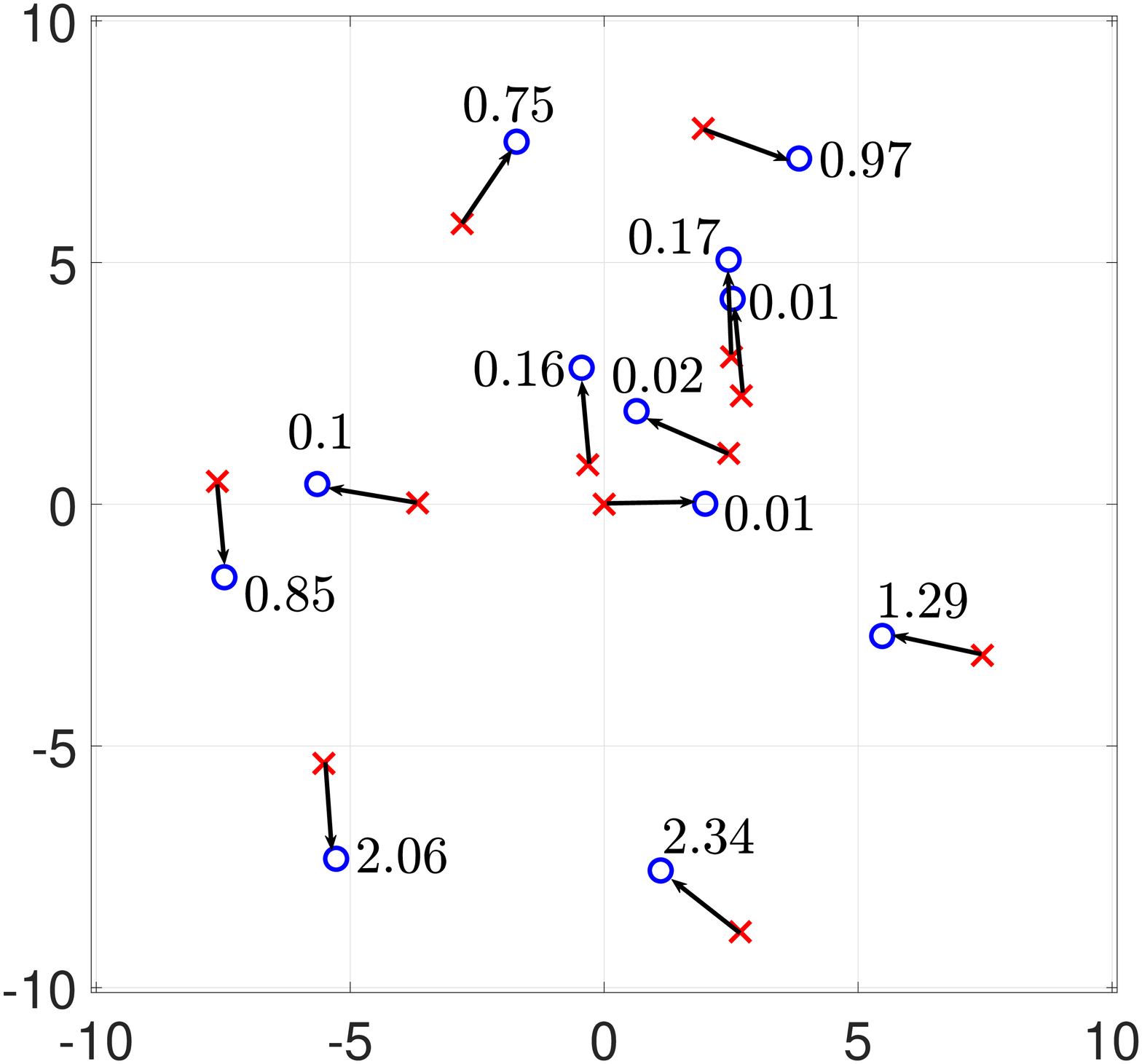}}
\caption{A realization of the Poisson bipolar network for $\lambda = 1/40$, $\alpha = 4$, and $R = 2$. `\textcolor{red}{$\times$}' denotes a transmitter while a circle `\textcolor{blue}{$\circ$}' denotes the associated receiver. In $(\mathrm{a})$, the number next to each link is a value of $P_{\rm s}$ (reliability) for that link for a deterministic SIR threshold of $\theta = 1$, while in $(\mathrm{b})$, the number next to each link is a value of the SIR threshold $T$ for that link such that the reliability is exactly $\nu$.}
\label{fig:dual}
\end{figure}

Fig.~\ref{fig:dual} visualizes Thm.~\ref{thm:dual} and reveals the rate-reliability trade-off. Fig.~\subref*{fig:ps_dist} shows what reliabilities links in a realization of the Poisson bipolar network achieve for a given SIR threshold $\theta$. For this realization, $3$ out of $12$ links achieve a reliability greater than the target reliability of $\nu = 0.9$. On the other hand, for the same realization, Fig.~\subref*{fig:theta_dist} shows what SIR threshold is set at each link to achieve exactly the target reliability. For those links that achieve a reliability greater than $\nu$ (see Fig.~\subref*{fig:ps_dist}), the SIR threshold can be set above the deterministic SIR threshold of $\theta = 1$ such that the links achieve exactly the target reliability (see Fig.~\subref*{fig:theta_dist}). For the links with smaller reliability than $\nu$, the SIR threshold needs to be lowered so that those links meet the target reliability. 

\subsection{Distribution of the SIR Threshold} 
\begin{theorem}
\label{thm:distr_T}
The random SIR threshold $T$ that guarantees a target reliability of $\nu = 1-\varepsilon$ is the solution of 
\begin{align}
\prod_{y\in \Phi} \frac{1}{1+ T R^{\alpha} \| y \|^{-\alpha}} = 1 -\varepsilon.
\label{eq:sir_thresh_eq}
\end{align}
\end{theorem}
\begin{IEEEproof}
The conditional success probability is given by
\begin{align}
P_{\rm s}(t) &= \mathbb{P}(\mathsf{SIR} > t \mid \Phi) \nonumber \\
&= \mathbb{P}\Bigg(\frac{hR^{-\alpha}}{\displaystyle \sum_{y \in \Phi}h_y\|y\|^{-\alpha}}> t \mid \Phi \Bigg)\nonumber \\
& \overset{(\mathrm{a})}{=} \mathbb{E}\Bigg(\exp\Bigg(-t R^{\alpha} \sum_{y \in \Phi}h_y\|y\|^{-\alpha} \Bigg) \mid \Phi\Bigg) \nonumber \\
&\overset{(\mathrm{b})}{=} \prod_{y\in \Phi} \frac{1}{1+ t R^{\alpha} \| y \|^{-\alpha}},
\end{align}
where step $(\mathrm{a})$ follows from averaging over the channel power gain $h$ of the desired link and step $(\mathrm{b})$ follows from averaging over the channel power gains $h_y$ of the interfering links.

Conditioned on $\Phi$, $T = t$ is the SIR threshold such that $\prod_{y\in \Phi} \frac{1}{1+ t R^{\alpha} \| y \|^{-\alpha}} = 1 -\varepsilon$. For a random $\Phi$, $T$ is a random variable. 
\end{IEEEproof}
\begin{remark}
\label{rem:gp}
Although in most cases, it is impossible to express $T$ in an exact closed-form, it can be computed numerically with ease.  Thanks to Thm.~\ref{thm:dual}, the distribution of $T$ can be exactly calculated from an already known exact expression of $\bar{F}_{P_{\rm s}(t)}(\nu)$ obtained via the Gil-Pelaez inversion~\cite{martin_meta_2016}. It can also be efficiently calculated by the binomial mixtures method~\cite{mh_binom}.
\end{remark}

We now discuss some cases of practical importance where we can express $T$ in a quasi-closed-form and calculate its distribution. For this purpose, we use the following lemma.
\begin{lemma}
\label{lem:lemma}
The relation between the geometric mean and the arithmetic mean gives rise to
\begin{align}
\left( \prod_{n = 1}^{k} 1+T R^{\alpha}R_n^{-\alpha} \right)^{1/k} \leq \frac{1}{k}\sum_{n = 1}^{k}(1+ T R^{\alpha}R_n^{-\alpha}).
\label{eq:gp_ap_inq}
\end{align}
\end{lemma}
In Lemma~\ref{lem:lemma}, the interference only from $k$ nearest interferers is considered, where $R_n$ is the $n$th interferer's distance from the typical receiver ({\em i.e.}, $R_1 \leq R_2 \leq \dotsc \leq R_k$).
\subsubsection{Ultrareliable regime ($\varepsilon \to 0$)}
The ultrareliable regime corresponds to the regime where the target outage probability goes to 0, {\em i.e.}, $\varepsilon \to 0$. For this regime, the following theorem calculates the distribution of $T$.
\begin{theorem}
\label{thm:ultrarel}
For $\varepsilon \to 0$,
\begin{align}
\bar{F}_T(t) \sim 1-\bar{F}_I\left(\frac{\varepsilon R^{-\alpha}}{t}\right),
\end{align}
where
\begin{align}
I \triangleq  \sum_{n = 1}^{\infty} R_n^{-\alpha}
\label{eq:intf_nofad}
\end{align}
denotes the interference without fading.
\end{theorem}
\begin{IEEEproof}
See the appendix.
\end{IEEEproof}
For $\alpha = 4$, $\bar{F}_I(\cdot)$ is available in a quasi-closed-form~\cite[Chapter 5]{martin_book}, which results in
\begin{align}
\bar{F}_T(t) \sim \mathrm{erfc}\left(\sqrt{\frac{t}{\varepsilon}}\frac{\pi^{3/2}\lambda R^2}{2}\right), \quad \varepsilon \to 0,
\label{eq:distr_thresh}
\end{align}
where $\mathrm{erfc}(\cdot)$ denotes the complementary error function. 
\begin{remark}
In \eqref{eq:distr_thresh}, the ratio of $t$ (rate) and $\varepsilon$ (reliability) highlights the rate-reliability trade-off.
\end{remark}
\subsubsection{Availability of partial (local) information about $\Phi$} In this case the knowledge of the locations of only a few nearest interferers is available. For this, the following theorem provides an approximate expression of the SIR threshold $T$.
\begin{theorem}
\label{thm:approx_T}
For any $\varepsilon \in (0, 1)$ and any $k \in \mathbb{N}$,
\begin{align}
T \geq \frac{k R^{-\alpha}\left(\left(\frac{1}{1-\varepsilon}\right)^{1/k} - 1\right)}{\sum_{n = 1}^{k} R_n^{-\alpha}}.
\label{eq:sir_thresh_inequality}
\end{align}
\end{theorem}
\begin{IEEEproof}
The proof follows directly from \eqref{eq:sir_thresh_eq} and \eqref{eq:gp_ap_inq}.
\end{IEEEproof}
\noindent For $k \to \infty$, the bound in \eqref{eq:sir_thresh_inequality} provides an approximation as
\begin{align}
T \approx \lim_{k \to \infty} \frac{k \left(\left(\frac{1}{1-\varepsilon}\right)^{1/k} - 1\right)R^{-\alpha}}{\sum_{n = 1}^{k} R_n^{-\alpha}} = \frac{\log (\frac{1}{1-\varepsilon})R^{-\alpha}}{I},
\label{eq:T_k_inf}
\end{align}
where $I$ is given by \eqref{eq:intf_nofad}. An approximate distribution of $T$ in a quasi-closed-form follows as
\begin{align}
\bar{F}_T(t) \approx \mathrm{erfc}\left(\sqrt{\frac{t}{\log (\frac{1}{1-\varepsilon})}}\frac{\pi^{3/2}\lambda R^2}{2}\right).
\label{eq:approx_T_dist}
\end{align}
\begin{remark}
By Thm.~\ref{thm:dual}, for a target reliability $\nu$, \eqref{eq:approx_T_dist} can be used to approximate $\bar{F}_{P_{\rm s}(t)}(\nu)$ as a function of $t$.
\end{remark}
%\begin{figure}
%\centering
%\includegraphics[scale=.45]{gp_ap_approx}
%\caption{Accuracy of the approximation in \eqref{eq:sir_thresh_inequality}: For the given parameters, the approximation with all interferers is quite accurate, which is expected for a high target reliability value due to the nature of \eqref{eq:sir_thresh_eq}.}
%\label{fig:gp_ap_approx}
%\end{figure} 

\begin{figure}
\centering
\includegraphics[width=7.8cm,height=6.3cm]{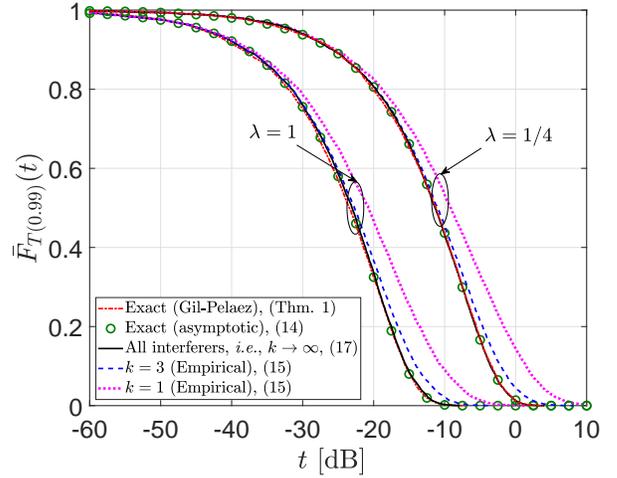}
\caption{Accuracy of the bound given by \eqref{eq:sir_thresh_inequality} for different values of densities.  The `Exact (Gil-Pelaez)' curve corresponds to the distribution of $T$ calculated using Thm.~\ref{thm:dual} and the exact expression of $\bar{F}_{P_{\rm s}(t)}(\nu)$ obtained by the Gil-Pelaez inversion, as stated in Remark~\ref{rem:gp}. The curves corresponding to the local information of $\Phi$, {\em i.e.,} $k = 1$ (nearest interferer) and $k = 3$ (three nearest interferers) denote the empirical distributions of $T$ obtained by treating the bound in \eqref{eq:sir_thresh_inequality} as an approximation. $\alpha = 4$ and $R  =1/2$.}
\label{fig:gp_ap_approx_diff_intf}
\end{figure} 

Fig.~\ref{fig:gp_ap_approx_diff_intf} shows that the empirical distribution of $T$ obtained using the bound given by \eqref{eq:sir_thresh_inequality} is quite tight for small values of the SIR threshold $t$. In fact, it is not necessary to have the entire information about the point process $\Phi$, {\em i.e.}, the locations of all interferers, to obtain a good approximation of $\bar{F}_T(t)$.

\subsection{Throughput Density}
The distribution of $T$ allows us to calculate the throughput density~\cite{sanket_globecom17}, which is defined as 
\begin{align}
S \triangleq \lambda \mathbb{E}(\log(1+T)P_{\rm{s}}(T)).
\label{eq:unrel_thro}
\end{align}
This throughput density metric takes into account all the links, which includes unreliable links for which $P_{\rm s} < 1-\varepsilon$. The second throughput metric, termed the {\em reliable} throughput density, considers only reliable links and is given by 
\begin{align}
S_{\rm{rel}} \triangleq  \lambda \mathbb{E}(\log(1+T)\boldsymbol{1}(P_{\mathrm{s}}(T)\geq 1-\varepsilon)),
\label{eq:thro2}
\end{align}
where $\boldsymbol{1}(\cdot)$ denotes the indicator function.

\subsubsection{Rate control approach} 
Since each link achieves the target reliability, we have $P_{\rm{s}}(T) = 1-\varepsilon$ and $\boldsymbol{1}(P_{\mathrm{s}}(T)\geq 1-\varepsilon) = 1$. It follows that
\begin{align}
S = \lambda(1-\varepsilon) \mathbb{E}(\log(1+T))
\end{align}
and
\begin{align}
S_{\rm{rel}} =  \lambda \mathbb{E}(\log(1+T)) =  \frac{S}{1-\varepsilon}.
\end{align}
For $\alpha = 4$, using \eqref{eq:approx_T_dist}, we can calculate $\mathbb{E}(\log(1+T))$, which gives rise to
\begin{align}
\mathbb{E}(\log(1+T)) &\approx \frac{1}{\sqrt{\pi}}\int_{0}^{\infty} u^{-1/2}e^{-u}\log\left(1+\frac{u}{C}\right)\mathrm{d}u\nonumber \\
&=-2C\:_2F_2([1, 1], [3/2, 2], C)  +\pi C\mathrm{erfi}(\sqrt{C})  \nonumber \\
&- (\gamma + 2 \log(2)+\log(C)),
\end{align}
where $C = -\frac{1}{\log(1-\varepsilon)}\frac{\pi^3 \lambda^2 R^4}{4}$, $_2F_2([\cdot, \cdot], [\cdot, \cdot], \cdot)$ is a hypergeometric function, $\mathrm{erfi}(\cdot)$ is the imaginary error function, and $\gamma = 0.57721$ is Euler's constant.
\begin{figure}
\centering
\includegraphics[width=8cm,height=6.5cm]{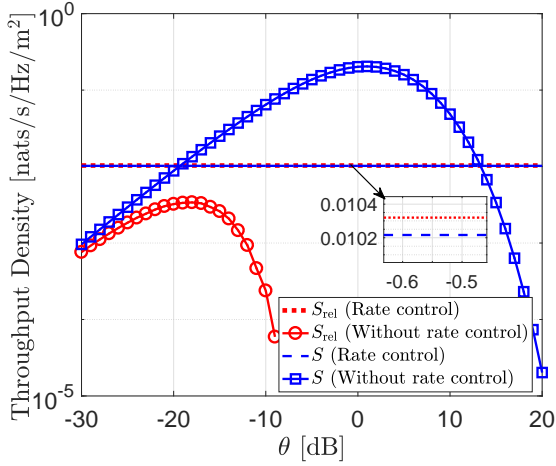}
\caption{Throughput densities against the deterministic SIR threshold $\theta$. $\lambda = 1$, $\alpha = 4$, $R  =1/2$, and $\varepsilon = 0.01$.}
\label{fig:thro_dens}
\end{figure}
\subsubsection{Deterministic SIR threshold approach}
The throughput densities are calculated by replacing the random $T$ in \eqref{eq:unrel_thro} and \eqref{eq:thro2} by the deterministic SIR threshold $\theta$. It follows that 
\begin{align}
S = \lambda\log(1+\theta)p_{\rm s}(\theta)
\end{align}
and
\begin{align}
S_{\rm{rel}} =  \lambda \log(1+\theta)F_{P_{\rm s}}(\theta, 1-\varepsilon),
\end{align}
where $p_{\rm s}(\theta) \triangleq \mathbb{E}(P_{\rm s}(\theta))$ is the standard success probability and $F_{P_{\rm s}}(\theta, 1-\varepsilon)  = \mathbb{E}(\boldsymbol{1}(P_{\mathrm{s}}(\theta)\geq 1-\varepsilon)) $ is the SIR MD. For the network model considered in this paper, from \cite{martin_meta_2016}, we have $p_{\rm s}(\theta) = \exp(-\lambda \pi R^2 \theta^{2/\alpha}\Gamma(1+2/\alpha)\Gamma(1-2/\alpha))$. $F_{P_{\rm s}}(\theta, 1-\varepsilon)$ can be calculated using the Gil-Pelaez inversion~\cite{martin_meta_2016}.

Fig.~\ref{fig:thro_dens} shows that when only reliable links are considered to calculate the throughput density, the rate control approach outperforms the deterministic SIR threshold ({\em i.e.}, without rate control) approach since all links are reliable in the former and only a fraction of links are reliable in the latter. When unreliable links are also considered, the trade-off between the rate and the reliability causes the rate control approach to perform better at low and high SIR thresholds. For the rest of the SIR threshold values, the rate control leads to smaller rates in an effort to make all links reliable and performs worse than the deterministic SIR threshold approach. The throughput densities for the rate control approach remain the same as they depend only on the average rate and the transmitter density.

\section{Conclusions}
For any wireless network modeled by a stationary and ergodic point process, we revealed that rate control is a facet of the SIR MD and showed its connection to the per-link reliability---another facet of the SIR MD. This connection permits the calculation of the rate distribution and highlights the rate-reliability trade-off. Since the rate control guarantees each link an arbitrary target reliability, it facilitates ultrareliable communication.\vspace*{1mm}

\appendix
\section*{Proof of Thm.~\ref{thm:ultrarel}}
The bound in \eqref{eq:gp_ap_inq} becomes exact as $\varepsilon \to 0$ because the term $T R^{\alpha} \| y \|^{-\alpha}$ in \eqref{eq:sir_thresh_eq} (hence the term $T R^{\alpha}R_n^{-\alpha}$ in \eqref{eq:gp_ap_inq}) approaches $0$ and hence the geometric mean approaches the arithmetic mean. Consequently, from \eqref{eq:sir_thresh_eq} and \eqref{eq:gp_ap_inq}, we have
\begin{align}
\lim_{k \to \infty}\frac{1}{k}\sum_{n = 1}^{k}(1+ TR^{\alpha}R_n^{-\alpha}) \sim \frac{1}{1-\varepsilon}, \quad \varepsilon \to 0,
\end{align}
which yields the exact expression of $T$ in the ultrareliable regime ({\em i.e.}, as $\varepsilon \to 0$) as
\begin{align}
T \sim \frac{\varepsilon R^{-\alpha}}{{\displaystyle \lim_{k \to \infty}}\sum_{n = 1}^{k} R_n^{-\alpha}}, \quad \varepsilon \to 0.
\end{align}
The exact distribution of $T$ follows as
\begin{align*}
\bar{F}_T(t) &= \mathbb{P}(T > t) \nonumber \\
&\sim 1- \mathbb{P}\left(I > \frac{\varepsilon R^{-\alpha}}{t}\right), \quad \varepsilon \to 0, \nonumber \\ 
&= 1-\bar{F}_I\left(\frac{\varepsilon R^{-\alpha}}{t}\right).
\end{align*}

\vspace*{1mm}

\bibliographystyle{ieeetr}

\end{document}